\documentclass[10pt,preprint2]{aastex}

\def \apj {ApJ}

\def \solphys {Solar Phys.}

\def \aap {A\&A}

\def \deg {$^{\rm o}$}

\newcommand{\citeN}[1]{\citeauthor{#1} (\citeyear{#1})}
\newcommand{\citeNP}[1]{\citeauthor{#1} \citeyear{#1}}

\newcommand{\FeI}{\ion{Fe}{1}}

\shortauthors{Socas-Navarro et al}
\shorttitle{Multi-Line Quiet Sun Spectro-Polarimetry at 5250 and 6302 \AA}

\begin{document}

\title{Multi-Line Spectro-Polarimetry of the Quiet Sun at 5250 and
  6302 \AA }

\author{H. Socas-Navarro, J. M. Borrero}
   	\affil{High Altitude Observatory, NCAR\thanks{The National Center
	for Atmospheric Research (NCAR) is sponsored by the National Science
	Foundation.}, 3080 Center Green Dr, Boulder, CO 80301, USA}
	\email{navarro@ucar.edu}

\author{A. Asensio Ramos, M. Collados, I. Dom\'\i nguez Cerde\~ na,
   	E. V. Khomenko, M. J. Mart\'\i nez Gonz\'
   	alez, V. Mart\'\i nez Pillet,
   	B. Ruiz Cobo, J. S\' anchez Almeida}
   	\affil{Instituto de Astrof\'\i sica de Canarias, Avda V\'\i a L\'
   	actea S/N, La Laguna 38205, Tenerife, Spain}

\date{}%

\begin{abstract}
The reliability of quiet Sun magnetic field diagnostics based on the
\ion{Fe}{1} lines at 6302 \AA \, has been questioned by recent
work. We present here the results of a thorough study of
high-resolution multi-line observations taken with the new
spectro-polarimeter SPINOR, comprising the 5250 and 6302 \AA \,
spectral domains. The observations were analyzed using several
inversion algorithms, including Milne-Eddington, LTE with 1 and 2
components, and MISMA codes. We find that the line-ratio technique
applied to the 5250~\AA \, lines is not sufficiently reliable to
provide a direct magnetic diagnostic in the presence of thermal
fluctuations and variable line broadening. In general, one needs to
resort to inversion algorithms, ideally with realistic
magneto-hydrodynamical constrains. When this is done, the 5250~\AA \,
lines do not seem to provide any significant advantage over those at
6302~\AA . In fact, our results point towards a better performance
with the latter (in the presence of turbulent line broadening). In any
case, for very weak flux concentrations, neither spectral region alone
provides sufficient constraints to fully disentangle the intrinsic
field strengths. Instead, we advocate for a combined analysis of both
spectral ranges, which yields a better determination of the quiet Sun
magnetic properties. Finally, we propose the use of two other
\ion{Fe}{1} lines (at 4122 and 9000~\AA ) with identical line opacities
that seem to work much better than the others.
\end{abstract}
   

\section{Introduction}
\label{sec:intro}

The empirical investigation of quiet Sun\footnote{In this work, we use the term
  ``quiet Sun'' to refer to the solar surface away from sunspots and
  active regions.}
magnetism is a very important
but extremely challenging problem. A large (probably dominant)
fraction of the solar magnetic flux resides in magnetic accumulations
outside active regions, forming network and inter-network patches
(e.g., \citeNP{SNSA02}). It is difficult to obtain conclusive observations
of these structures, mainly because of two reasons. First, the
size of the magnetic concentrations is much smaller than the spatial
resolution capability of modern spectro-polarimetric
instrumentation. Estimates obtained with inversion codes yield typical
values for the filling factor of the resolution element between
$\sim$1\% and 30\% . The interpretation of the polarization signal
becomes non-trivial in these conditions and one needs to make
use of detailed inversion codes to infer the magnetic field in the
atmosphere. Second, the observed signals are extremely weak (typically
below $\sim$1\% of the average continuum intensity), demanding
both high sensitivity and high resolution. Linear polarization is
rarely observed in visible lines, so one is usually left with
  Stokes~$I$ and~$V$ alone. 

\citeN{S73} proposed to use the pair of \ion{Fe}{1} lines at 5247
and~5250~\AA \, which have very similar excitation potentials and
oscillator strengths (and, therefore, very similar opacities) but
  different Land\'e factors, to
determine the intrinsic field strength directly from the Stokes~$V$
line ratio. That work led to the subsequent popularization of this
spectral region for further studies of unresolved solar magnetic
structures. Later, the pair of \ion{Fe}{1} lines at 6302~\AA \, became
the primary target of the Advanced Stokes Polarimeter (ASP,
\citeNP{ELT+92}), mainly due to their lower sensitivity to temperature
fluctuations. The success of the ASP has contributed largely to the
currently widespread use of the 6302~\AA \, lines by the solar
community.

Recent advances in infrared spectro-polarimetric instrumentation now
permit the routine observation of another very interesting pair of
\ion{Fe}{1} lines, namely those at 15648 and 15653~\AA \, (hereafter,
the 1.56~$\mu$m lines). Examples are the works of \citeN{LR99};
\citeN{KCS+03}. The large Land\' e factors of these lines,
combined with their very long wavelengths, result in an extraordinary
Zeeman sensitivity. Their Stokes~$V$ profiles exhibit patterns where
the $\sigma$-components are completely split for fields stronger than
$\sim$400~G at typical photospheric conditions. They also produce
  stronger linear polarization. On the downside,
this spectral range is accesible to very few
polarimeters. Furthermore, the 1.56~$\mu$m lines are rather weak in
comparison with the above-mentioned visible lines.

Unfortunately, the picture revealed by the new infrared data often
differs drastically from what was being inferred from the 6302~\AA \,
observations (e.g., \citeNP{LR99}; \citeNP{KCS+03}; \citeNP{SNSA02};
\citeNP{SNL04}; \citeNP{DCSAK03}), particularly in the
  inter-network. \citeN{SNSA03} proposed that the 
discrepancy in the field strengths inferred from the visible and
infrared lines may be explained by magnetic inhomogeneities within the
resolution element (typically 1\arcsec). If multiple field strengths
coexist in the observed pixel, then the infrared lines will be more
sensitive to the weaker fields of the distribution whereas the visible
lines will provide information on the stronger fields (see also the
discussion about polarimetric signal increase in the 1.56 $\mu$m lines
with weakening fields in \citeNP{SAL00}). This conjecture has been
tested recently by \citeN{DCSAK06} who modeled simultaenous
observations of visible and infrared lines using unresolved magnetic
inhomogeneities.

A recent paper describing numerical simulations by \citeN{MGCRC06}
casts some doubts on the results obtained using the 6302~\AA \,
lines. Our motivation for the present work is to resolve this issue by
observing simultaneously the quiet Sun at 5250 and 6302~\AA . We know
that unresolved magnetic structure might result in different field
determinations in the visible and the infrared, but the lines analyzed
in this work are close enough in wavelengths and Zeeman sensitivities
that one would expect to obtain the same results for both spectral
regions.

\section{Methodology}
\label{sec:method}

Initially, our goal was to observe simultaneously at 5250 and
  6302~\AA \ because we expected the 5250~\AA \ lines to be a very
  robust indicator of intrinsic field strength, which we could then
  use to test under what conditions the 6302~\AA \ lines are also
  robust. Unfortunately, as we show below, it turns out that in most
  practical situations the 5250~\AA \ lines are not more robust than
  the 6302~\AA \ pair. This left us without a generally valid
  reference frame against which to test the 6302~\AA \ lines. We then
  decided to employ a different approach for our study, namely to
  analyze the uniqueness of the solution obtained when we invert the
  lines and how the solutions derived for both pairs of lines compare
  to each other. In doing so, there are some subtleties that need to
  be taken into consideration.

   An inversion technique necessarily resorts on a number of physical
    assumptions on the solar atmosphere in which the lines are formed
    and the (in general polarized) radiative transfer. This implies
    that the conclusions obtained from applying a particular inversion
    code to our data may be biased by the modeling implicit in the
    inversion. Therefore, a rigorous study requires the analysis of
    solutions from a wide variety of inversion procedures. Ideally,
    one would like to cover at least the most frequently employed
    algorithms. For our purposes here we have chosen four of the most
    popular codes, spanning a wide range of model complexity. They are
    described in some detail in section~\ref{sec:obs} below.

When dealing with Stokes inversion codes, there are two very distinct
    problems that the user needs to be aware of. First, it may happen
    that multiple different solutions provide satisfactory
    fits to our observations. This problem is of a very fundamental
    nature. It is not a problem with the inversion algorithm but with
    the observables themselves. Simply put, they do not carry
    sufficient information to discriminate among those particular
    solutions. The only way around this problem is to either supply
    additional observables (e.g., more spectral lines) or to restrict
    the allowed range of solutions by incorporating sensible
    constraints in the physical model employed by the inversion. A
    second problem arises when the solution obtained does not fit the
    observed profiles satisfactorily. This can happen because the
    physical constraints in the inversion code are too stringent (and
    thus no good solution exists within the allowed range of
    parameters), or simply because the algorithm happened to stop at a
    secondary minimum. This latter problem is not essential because
    one can always discard ``bad'' solutions (i.e., those that result
    in a bad fit to the data) and simply try again with a different
    initialization. In this work we are interested in exploring the
    robustness of the observables in as much as they relate to the
    former problem, i.e. the underlying uniqueness of the solution. We
    do this by performing a large number of inversions with random
    initializations and analyzing the dependence of the solutions with
    the merit function $\chi^2$ (defined below).

    Ideally, one would like to see that for small values of $\chi^2$,
    all the solutions are clustered around a central value with a
    small spread (the behavior for large $\chi^2$ is not very relevant
    for our purposes here). This should happen regardless of the
    inversion code employed and the pair of lines analyzed. If that
    were the case we could conclude that our observables are truly
    robust. Otherwise, one needs to be careful when analyzing data
    corresponding to that particular scenario.

\section{Observations and Data Reduction}
\label{sec:obs}

The observations used in this work were obtained during an observing
run in March 2006 with the Spectro-Polarimeter for Infrared and Optical
Regions (SPINOR, see \citeNP{SNEP+06}), attached to the Dunn Solar
Telescope (DST) at the Sacramento Peak Observatory (Sunspot, NM USA,
operated by the National Solar Observatory). SPINOR allows for the
simultaneous observation of multiple spectral domains with nearly
optimal polarimetric efficiency over a broad range of wavelengths.

The high-order adaptive optics system of the DST (\citeNP{RHR+03}) was
employed for image stabilization and to correct for atmospheric
turbulence. This allowed us to attain sub-arcsecond spatial resolution
during some periods of good seeing. 

The observing campaign was originally devised to obtain as much
information as possible on the unresolved properties of the quiet Sun
magnetic fields. In addition to the 5250 and 6302~\AA \, domains
discussed here, we also observed the \ion{Mn}{1} line sensitive to
hyperfine structure effects (\citeNP{LATC02}) at 5537~\AA \, and the
\FeI \ line pair at
1.56~$\mu$m.

SPINOR was operated in a configuration with four different detectors
which were available at the time of observations (see
Table~\ref{tab:detectors}): The Rockwell TCM 8600 infrared camera,
with a format of 1024$\times$1024 pixels, was observing the
1.56~$\mu$m region. The SARNOFF CAM1M100 of 1024$\times$512 pixels was
used at 5250~\AA . Finally, the two dual TI TC245 cameras of
256$\times$256 pixels (the original detectors of the Advanced Stokes
Polarimeter) were set to observe at 5537 and 6302~\AA . Unfortunately,
we encountered some issues during the reduction of the 5537~\AA \, and
1.56~$\mu$m data and it is unclear at this point whether or very not they
are usable. Therefore, in the remainder of this paper we shall focus on
the analysis of the data taken at 5250 and 6302~\AA . The spectral
resolutions quoted in the table are estimated as the quadratic sum of
the spectrograph slit size, diffraction limit and pixel sampling.

In order to have good spectrograph efficiency at all four wavelengths
simultaneously we employed the 308.57~line~mm$^{-1}$ grating (blaze
angle 52$^{\rm o}$), at the expense of obtaining a relatively low
dispersion and spectral resolution (see Table~\ref{tab:detectors} for
details).

\begin{deluxetable}{lccccc}
\tablewidth{0pt}
\tablecaption{SPINOR detector configuration
\label{tab:detectors}}
\tablehead{Camera & Wavelength & Spectral resolution & Spectral
  sampling & Usable range & Field of view \\
                  &  (\AA )    &  (m\AA )  & (m\AA ) & (\AA ) &  along
  the slit (\arcsec)}
\startdata
ROCKWELL   & 15650 &   190  &  150  &  150  &  187 \\
SARNOFF    & 5250  &   53  &  31  &  15  &  145 \\
TI TC245           & 5537  &   40  &  29  &  5  &  95 \\
TI TC245           & 6302  & 47  &  24  &  6  &  95 \\
\enddata
\end{deluxetable}

Standard flatfield and bias correction proceduress were applied to the
images. Subsequent processing included the removal of spectrum
curvature and the alignment of both polarized beams, using for this
purpose a pair of hairlines inserted across the slit. Calibration
operations were performed to determine the SPINOR polarimetric
response matrix by means of calibration optics located at the
telescope beam exit port. In this manner we can decontaminate the
datasets from instrumental polarization introduced by the
polarimeter. Finally, it is also important to consider the
contamination introduced by the telescope. To this aim we obtained
telescope calibration data with an array of linear polarizers situated
over the DST entrance window. By rotating these polarizers to
different angles, it is possible to feed light in known polarization
states into the system. A cross-dispersing prism was placed in front
of one of the detectors, allowing us to obtain calibration data
simultaneously across the entire visible spectrum. Details on the
procedure may be found in \citeN{SNEP+06}.

In this paper we focus on two scan operations near disk center, one
over a relatively large pore (at solar heliocentric coordinates
longitude -25.40, latitude -3.68) and the other of a quiet region
(coordinates longitude 0.01\deg , latitude -7.14\deg ). The pore map was taken
with rather low spatial resolution ($\sim$1.5\arcsec) but exhibits a
large range of polarization signal amplitudes. The quiet map, on the
other hand, has very good spatial resolution ($\sim$0.6\arcsec) but
the polarization signals are much weaker.  The noise level, measured
as the standard deviation of the polarization signal in continuum
regions, is approximately 7$\times$10$^{-4}$ and 5$\times$10$^{-4}$
times the average quiet Sun continuum intensity at 5250 and 6302~\AA ,
respectively.

We used several different inversion codes for the various tests
presented here, namely: SIR (Stokes Inversion based on Response
functions, \citeNP{RCdTI92}); MELANIE (Milne-Eddington Line Analysis
using a Numerical Inversion Engine) and LILIA (LTE Inversion based on
the Lorien Iterative Algorithm, \citeNP{SN01a}); and MISMA
(MIcro-Structured Magnetic Atmosphere, \citeNP{SA97}). The simplest of
these algorithms is MELANIE, which implements a Milne-Eddington type
of inversion similar to that of \citeN{SL87}. The free parameters
considered include a constant along the line of sight magnetic field
vector, magnetic filling factor, line-of-sight velocity and several
spectral line parameters that represent the thermal properties of the
atmosphere (Doppler width $\Delta \lambda_D$, line-to-continuum
opacity ratio $\eta_0$, source function $S$ and damping $a$). The
\ion{Fe}{1} lines at 6302~\AA \, belong to the same multiplet and
their $\eta_0$ are related by a constant factor. Assuming that the
formation height is similar enough for both lines we can also consider
that they have the same $\Delta \lambda_D$, $S$ and $a$. In this
manner the same set of free parameters can be used to fit both
lines. In the case of the 5250~\AA \, lines we only invert the
\ion{Fe}{1} pair and assume that both lines have identical opacities
$\eta_0$.

SIR considers a model atmosphere defined by the depth stratification
of variables such as temperature, pressure, magnetic field vector,
line-of-sight velocity and microturbulence. Atomic populations are
computed assuming LTE for the various lines involved, making it
possible to fit observations of lines from multiple chemical elements
with a single model atmosphere that is common to all of them. Unlike
MELANIE, one can produce line asymmetries by incorporating gradients
with height of the velocity and other parameters. LILIA is a different
implementation of the SIR algorithm. It works very similarly with some
practical differences that are not necessary to discuss here.

MISMA is another LTE code but has the capability to consider three
atmospheric components (two magnetic and one non-magnetic) that are
interlaced on spatial scales smaller than the photon mean free
path. Perhaps the most interesting feature of this code for our
purposes here is that it implements a number of magneto-hydrodynamic
(MHD) constrains, such as momentum, as well as mass and magnetic flux
conservation. In this manner it is possible to derive the full
vertical stratification of the model atmosphere from a limited number
of free parameters (e.g., the magnetic field and the velocity at the
base of the atmosphere).

In all of the inversions presented here we employed the same set of
atomic line parameters, which are listed in Table~\ref{tab:atomic}.

\begin{deluxetable}{cccccc}
\tablewidth{0pt}
\tablecaption{Atomic line data
\label{tab:atomic}}
\tablehead{    & Wavelength & Excit. Potential & $\log(gf)$
  & Lower level & Upper level \\
           Ion &  (\AA)     &    (eV)          &          
 & configuration & configuration }
\startdata
\ion{Fe}{1} & 4121.8020 & 2.832  &  -1.300  & $^3$P$_2$ & $^3$F$_3$\\
\ion{Cr}{2} & 5246.7680 & 3.714  &  -2.630  & $^4$P$_{1/2}$ & $^4$P$_{3/2}$\\
\ion{Fe}{1} & 5247.0504 & 0.087  &  -4.946  & $^5$D$_2$ & $^7$D$_3$\\
\ion{Ti}{1} & 5247.2870 & 2.103  &  -0.927  & $^5$F$_3$ & $^5$F$_2$\\
\ion{Cr}{1} & 5247.5660 & 0.961  &  -1.640  & $^5$D$_0$ & $^5$P$_1$\\
\ion{Fe}{1} & 5250.2089 & 0.121  &  -4.938  & $^5$D$_0$ & $^7$D$_1$\\
\ion{Fe}{1} & 5250.6460 & 2.198  &  -2.181  & $^5$P$_2$ & $^5$P$_3$\\
\ion{Fe}{1} & 6301.5012 & 3.654  &  -0.718  & $^5$P$_2$ & $^5$D$_2$\\
\ion{Fe}{1} & 6302.4916 & 3.686  &  -1.235  & $^5$P$_1$ & $^5$D$_0$\\
\ion{Fe}{1} & 8999.5600 & 2.832  &  -1.300  & $^3$P$_2$ & $^3$P$_2$\\
\enddata
\end{deluxetable}

\section{Results}
\label{sec:results}

\begin{figure*}
\epsscale{2}
\plotone{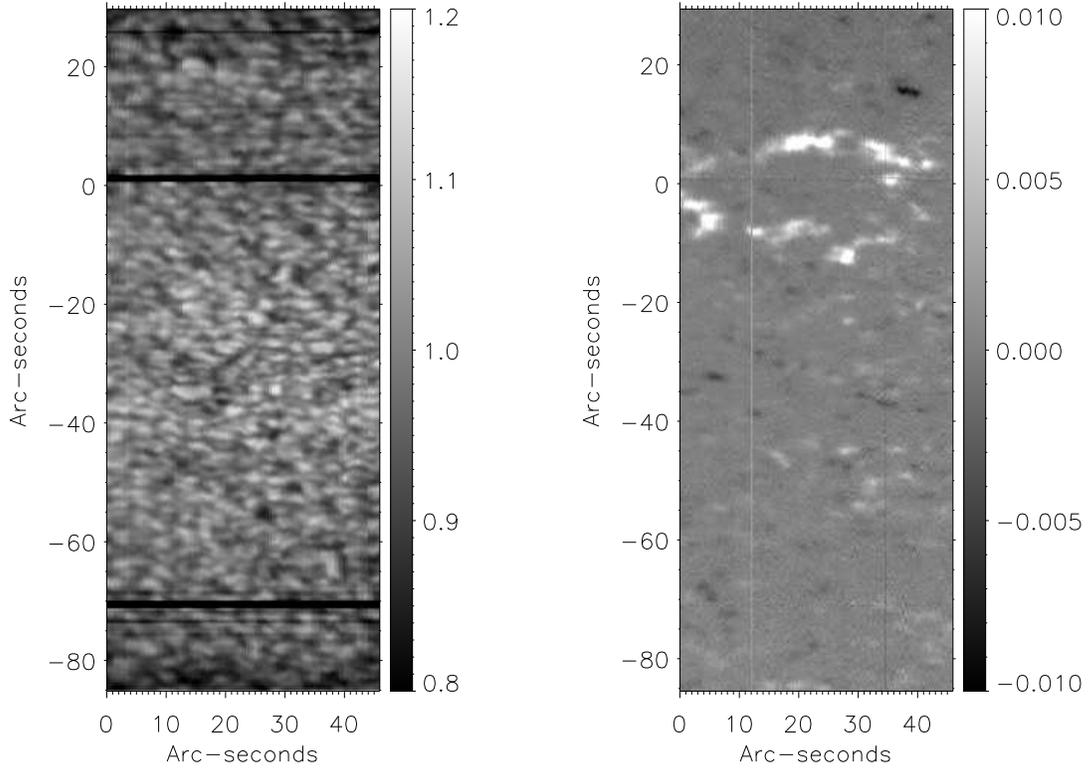}
\caption{Continuum intensity map (left) and reconstructed magnetogram
  (right) of the quiet Sun scan near 5250~\AA  . The magnetogram shows
  the integrated Stokes~$V$ signal over a narrow bandwidth on the red
  lobe of the \ion{Fe}{1} 5250.2~\AA \, line. In both panels, values
  are referred to the average quiet Sun intensity. The angular scale
  on the vertical axis measures the distance from the upper hairline.
\label{fig:map1}
}
\end{figure*}

\begin{figure*}
\epsscale{2}
\plotone{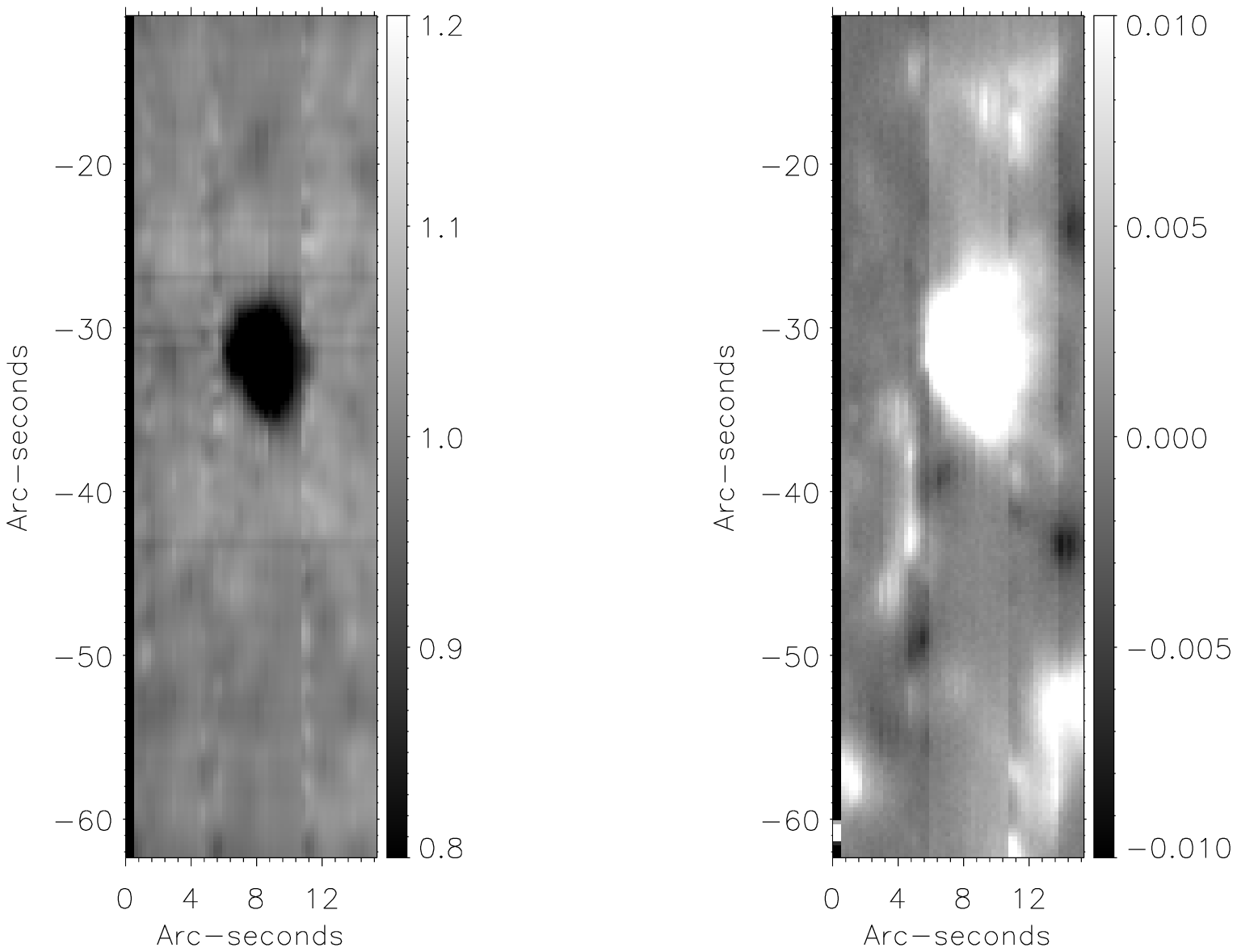}
\caption{Continuum intensity map (left) and reconstructed magnetogram
  (right) of the pore scan near 5250~\AA  . The magnetogram shows
  the integrated Stokes~$V$ signal over a narrow bandwidth on the red
  lobe of the \ion{Fe}{1} 5250.2~\AA \, line. In both panels, values
  are referred to the average quiet Sun intensity. The angular scale
  on the vertical axis measures the distance from the upper hairline.
\label{fig:map2}
}
\end{figure*}

Figures~\ref{fig:map1} and~\ref{fig:map2} show continuum maps and
reconstructed magnetograms of the pore and quiet Sun scans in the
5250~\AA \, region. Notice the much higher spatial resolution in the
quiet Sun observation (Fig~\ref{fig:map1}). Similar maps, but with a
somewhat smaller field of views, can be produced at 6302~\AA . 

\begin{figure*}
\epsscale{2}
\plotone{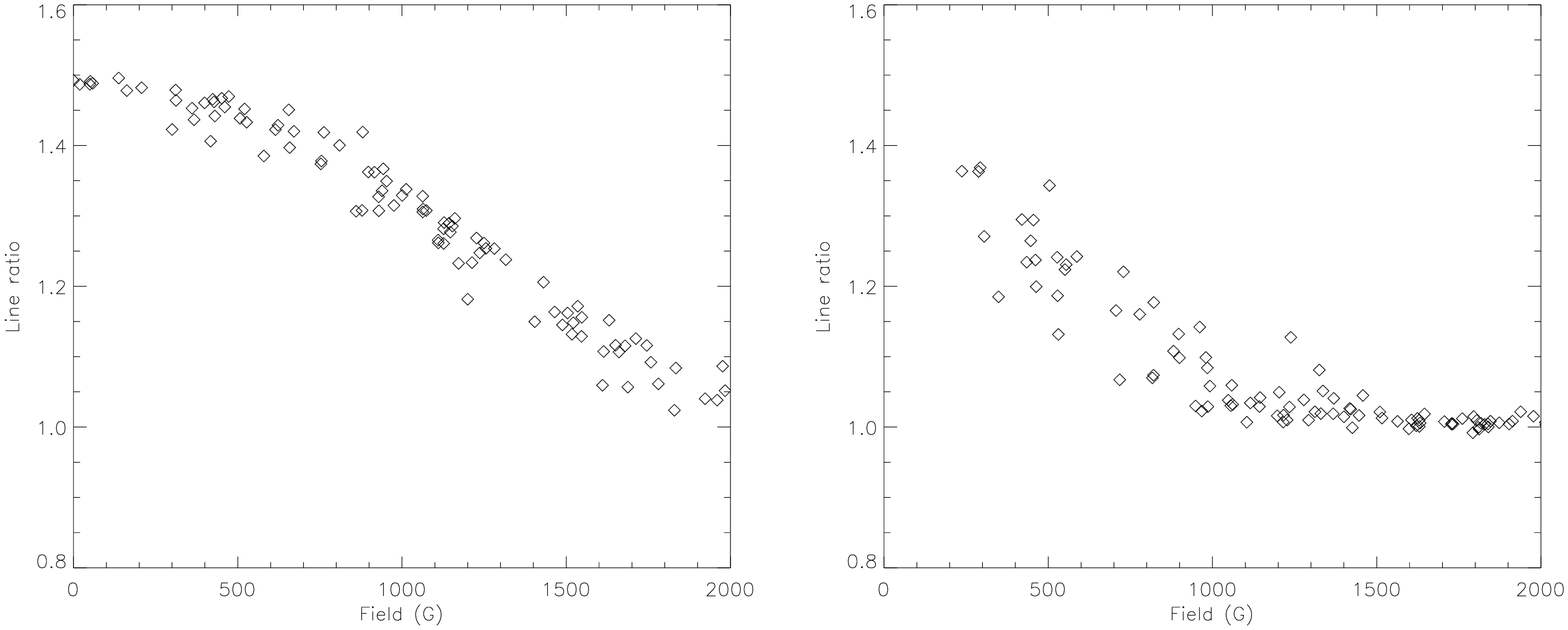}
\caption{Line ratio calibration using synthetic Stokes profiles
  of \FeI \ 5250~\AA \ 
  emergent from the HSRA model atmosphere after adding random
  temperature perturbations and magnetic fields. Left: Macroturbulence
  was kept fixed at 3~km~s$^{-1}$. Right: Random macroturbulence
  varies between 0 and 1~km~s$^{-1}$.
\label{fig:calib}
}
\end{figure*}

The first natural step in the analysis of these observations, before
even considering any inversions, is to calculate the Stokes~$V$
amplitude ratio of the \ion{Fe}{1} line 5250.2 to 5247.0~\AA \,
(hereafter, the line ratio). One would expect to obtain a rough idea
of the intrinsic magnetic field strength from this value alone.  A
simple calibration was derived by taking the thermal stratification of
the Harvard-Smithsonian Reference Atmosphere (HSRA, see
\citeNP{GNK+71}) and adding random (depth-independent) Gaussian
temperature perturbations with an amplitude (standard deviation) of
$\pm$300~K, different magnetic field strengths and a fixed
macroturbulence of 3~km~s$^{-1}$ (this value corresponds roughly to
our spectral resolution). Line-of-sight gradients of temperature,
field strength and velocity are also included.  Figure~\ref{fig:calib}
shows the line ratio obtained from the synthetic profiles as a
function of the magnetic field employed to synthesize them. The line
ratio is computed simply as the peak-to-peak amplitude ratio of the
Stokes~$V$ profiles. Notice that when the same experiment is carried
out with a variable macroturbulent velocity the scatter
increases considerably, even for relatively small values of up to
1~km~s$^{-1}$ (right panel). The syntheses of Fig~\ref{fig:calib}
consider the partial blends of all 6 lines in the 5250~\AA \, spectral
range.

\begin{figure*}
\epsscale{2}
\plotone{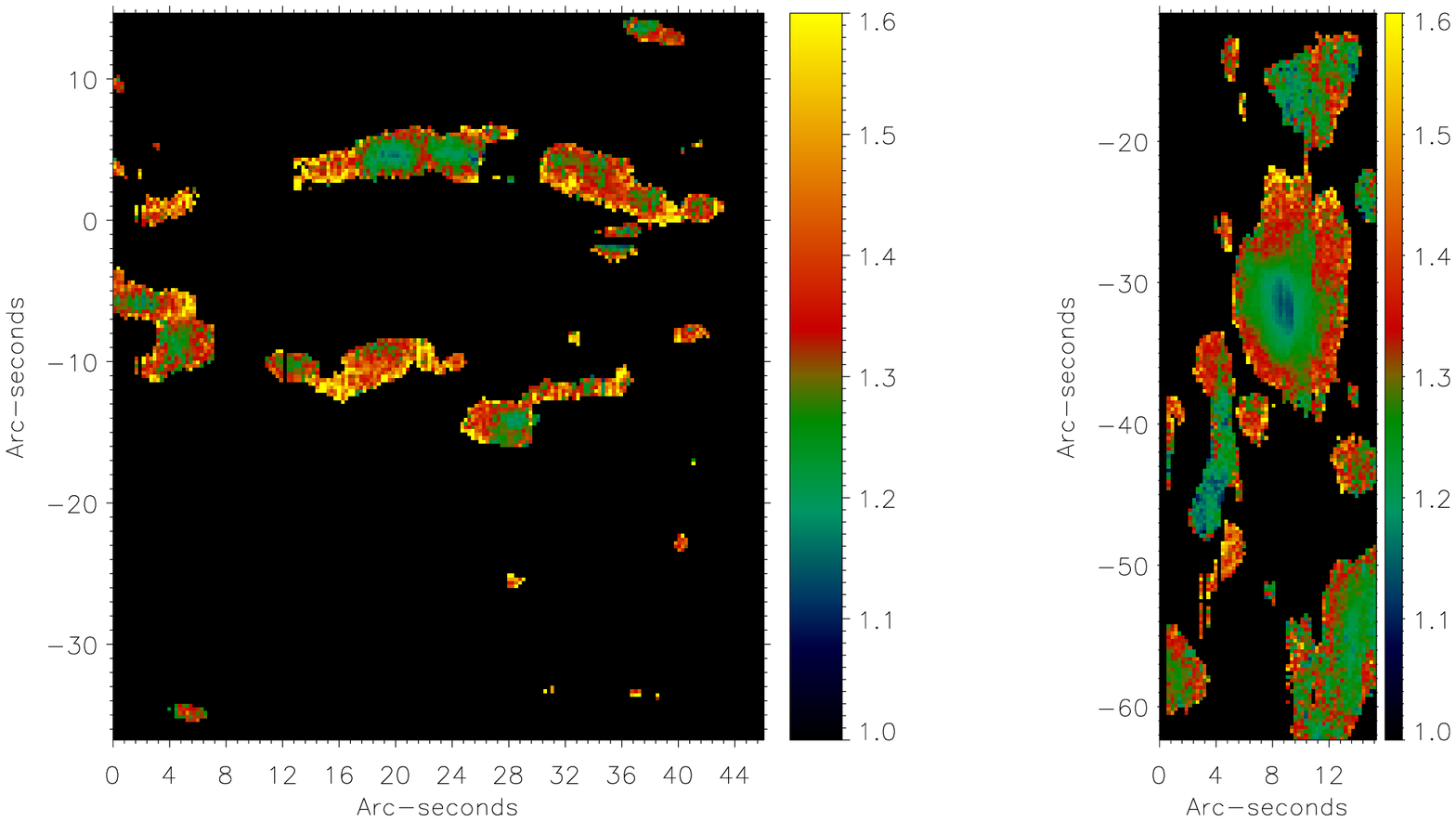}
\caption{Amplitude ratio of Stokes~$V$ in \ion{Fe}{1} 5250.2 to
  \ion{Fe}{1} 5247.0~\AA . Left: Network patches observed in the quiet
  Sun scan. Right: A small pore (centered on coordinates [10,-30]
  approximately) and surrounding plage. Ratios close to 1 are
  indicative of a strong field of nearly 2~kG. Ratios close to 1.6
  correspond to the weak-field (up to $\sim$500~G) regime. Black areas
  exhibit a circular polarization amplitude smaller than 1\% and have
  not been inverted.
\label{fig:mapratios}
}
\end{figure*}

Figure~\ref{fig:mapratios} shows the observed line ratios for both
scans. According to our calibration (see above), ratios close to 1
indicate strong fields of nearly (at least) $\sim$2~kG, whereas
larger values would suggest the presence of weaker fields, down to the
weak-field saturation regime at (at most) $\sim$500~G corresponding to
a ratio of 1.5. Figure~\ref{fig:mapratios} is somewhat disconcerting
at first sight. The pore exhibits the expected behavior with strong
$\sim$2~kG fields at the center that decrease gradually towards the
outer boundaries until it becomes weak. The network and plage patches,
on the other hand, contain relatively large areas with high ratios of
1.4 and even 1.6 at some locations. This is in sharp contrast with the
strong fields ($\sim$1.5~kG) that one would expect in network and
plage regions (e.g., \citeNP{SSH+84}; \citeNP{SAL00}; \citeNP{BRRCC00}).

\begin{figure}
\epsscale{1}
\plotone{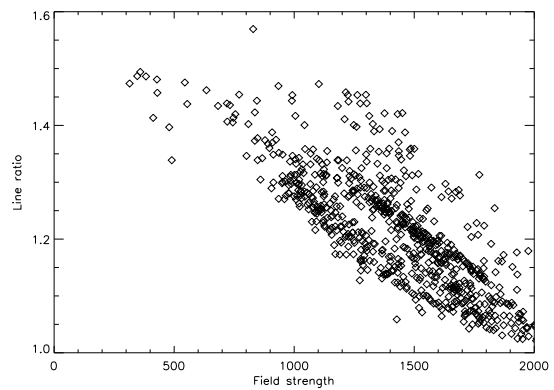}
\caption{Line ratio calibration using synthetic Stokes profiles
  obtained in the inversion of the pore.
\label{fig:calibpore}
}
\end{figure}

In view of these results we carried out inversions of the Stokes $I$
and $V$
profiles of the spectral lines in the 5250~\AA \, region emergent from
the pore. We used the code LILIA considering a single magnetic
component with a constant magnetic field embedded in a (fixed)
non-magnetic background, taken to be the average quiet Sun. The
resulting field-ratio scatter plot is shown in
Figure~\ref{fig:calibpore}. Note that the scatter in this case is much
larger than that obtained with the HSRA calibration above. It is
important to point out that the line ratio depicted in the figure is
that measured on the synthetic profiles. Therefore, the scatter cannot
be ascribed to inaccuracies of the inversion. As a verification test
we picked one of the models with kG fields that produced a line ratio
of $\simeq$1.4 and synthesized the emergent profiles with a different
code (SIR), obtaining the same ratio. We therefore conclude that, when
the field is strong, many different atmospheres are able to produce
similar line ratios if realistic thermodynamic fluctuations and
turbulence are allowed in the model. For weak fields, the ratio tends
to a value of $\simeq$1.5 without much fluctuations.

\begin{figure}
\epsscale{1}
\plotone{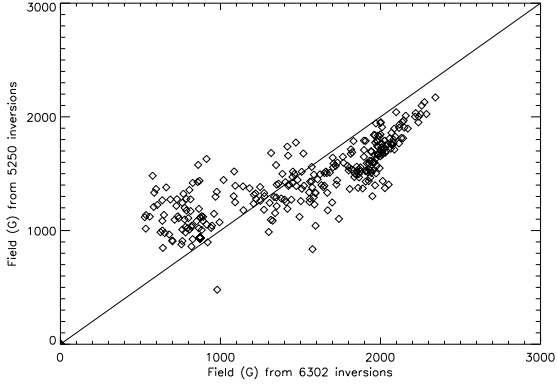}
\caption{Intrinsic field strengths determined from inverting the lines
  at 5250 (ordinates) and 6302 (abscissa) observed in the pore. The
  solid line shows the diagonal of the plot.
\label{fig:porefields}
}
\end{figure}

\begin{figure}
\epsscale{1}
\plotone{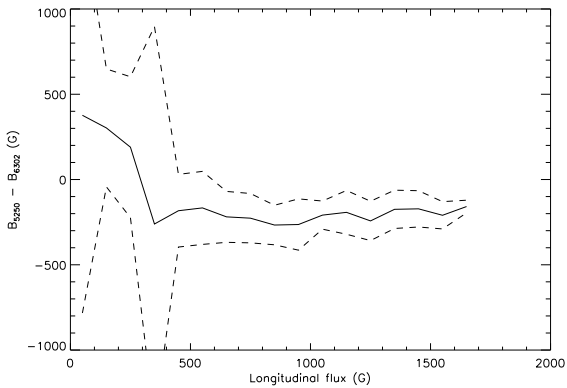}
\caption{Differences in intrinsic field strengths measured at 6302 and
  5250~\AA \, as a function of the average longitudinal magnetic flux
  density. The solid line represents the median value over 100~G
  bins. The dashed lines represent the standard deviation of the
  points in each bin.
\label{fig:poreerrors}
}
\end{figure}

Accepting then that we could no longer rely on the line ratio of
\ion{Fe}{1} 5247.0 and 5250.2~\AA \, as an independent reference to
verify the magnetic fields obtained with the 6302~\AA \, lines, we
considered the result of inverting each spectral region separately.
Figure~\ref{fig:porefields} depicts the scatter plot obtained. At the
center of the pore, where we have the strongest fields (right-hand
side of the figure), there is a very good correlation between the
results of both measurements. However, those points lay systematically
below the diagonal of the plot. This may be explained by the different
``formation heights'' of the lines. The \ion{Fe}{1} lines at 5250~\AA
\, generally form somewhat higher than those at 6302~\AA . If the
field strength decreases with height, one would expect to retrieve a
slightly lower field strength when using 5250~\AA . Unfortunately, the
correlation breaks down for the weaker fields. In
Figure~\ref{fig:poreerrors} we can see that both sets of lines yield
approximately the same field strengths (with a slightly lower values
for the 5250 inversions, as discussed above) for longitudinal fluxes
above some $\sim$500~G. Below this limit our diagnosis is probably
not sufficiently robust.

If instead of the intrinsic field we consider the longitudinal
magnetic flux, we obtain a fairly good agreement between both spectral
regions. The Milne-Eddington inversions with MELANIE yield a Pearsons
correlation coefficient of 0.89. In principle, the agreement is
somewhat worse for the LILIA inversions, with a correlation
coefficient of 0.60. However, we have found that this is due to a few
outlayer points. Removing them results in a correlation coefficient of
0.82.

\begin{figure}
\epsscale{1}
\plotone{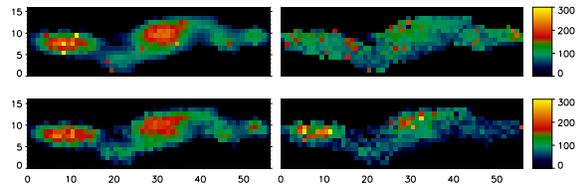}
\caption{Longitudinal flux density from the inversions of a network
  patch at 6302 (left) and 5250~\AA \, (right). Upper panels:
  Inversions with the Milne-Eddington code MELANIE. Lower panels:
  Inversions with the LTE code LILIA. Spatial scales are in pixels.
\label{fig:fluxes}
}
\end{figure}

The situation becomes more complicated in the network. The results of
inverting a network patch with MELANIE and LILIA can be seen in
Figure~\ref{fig:fluxes}. The inversions with MELANIE (upper panels) do
not agree very well with each other. The correlation coefficient is
only 0.23. The 5250~\AA \, map (upper right panel in the figure) looks
considerably more noisy and rather homogeneous, compared to the
6302~\AA \, map (upper left). The LILIA inversions (lower panels)
exhibit somewhat better agreement (correlation is 0.65), but again
look very noisy at 5250~\AA . 

\begin{figure}
\epsscale{1}
\plotone{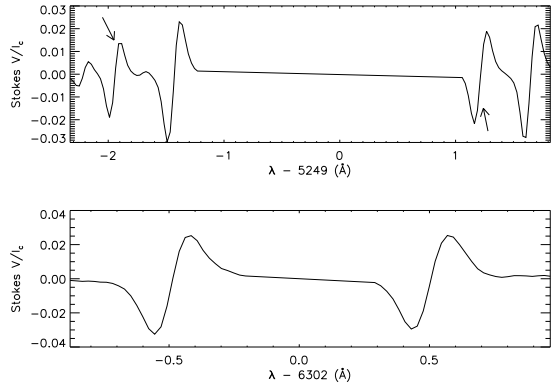}
\caption{Average observed Stokes~$V$ profiles in a network patch
  (located approximately around coordinates [20,7] in
  Figure~\ref{fig:map1}). 
Ordinate values are related to the average quiet Sun
  continuum intensity. The 5250 line ratio (the relevant \ion{Fe}{1}
  lines are marked with arrows) is 1.21.
\label{fig:NWa}
}
\end{figure}

\begin{figure}
\epsscale{1}
\plotone{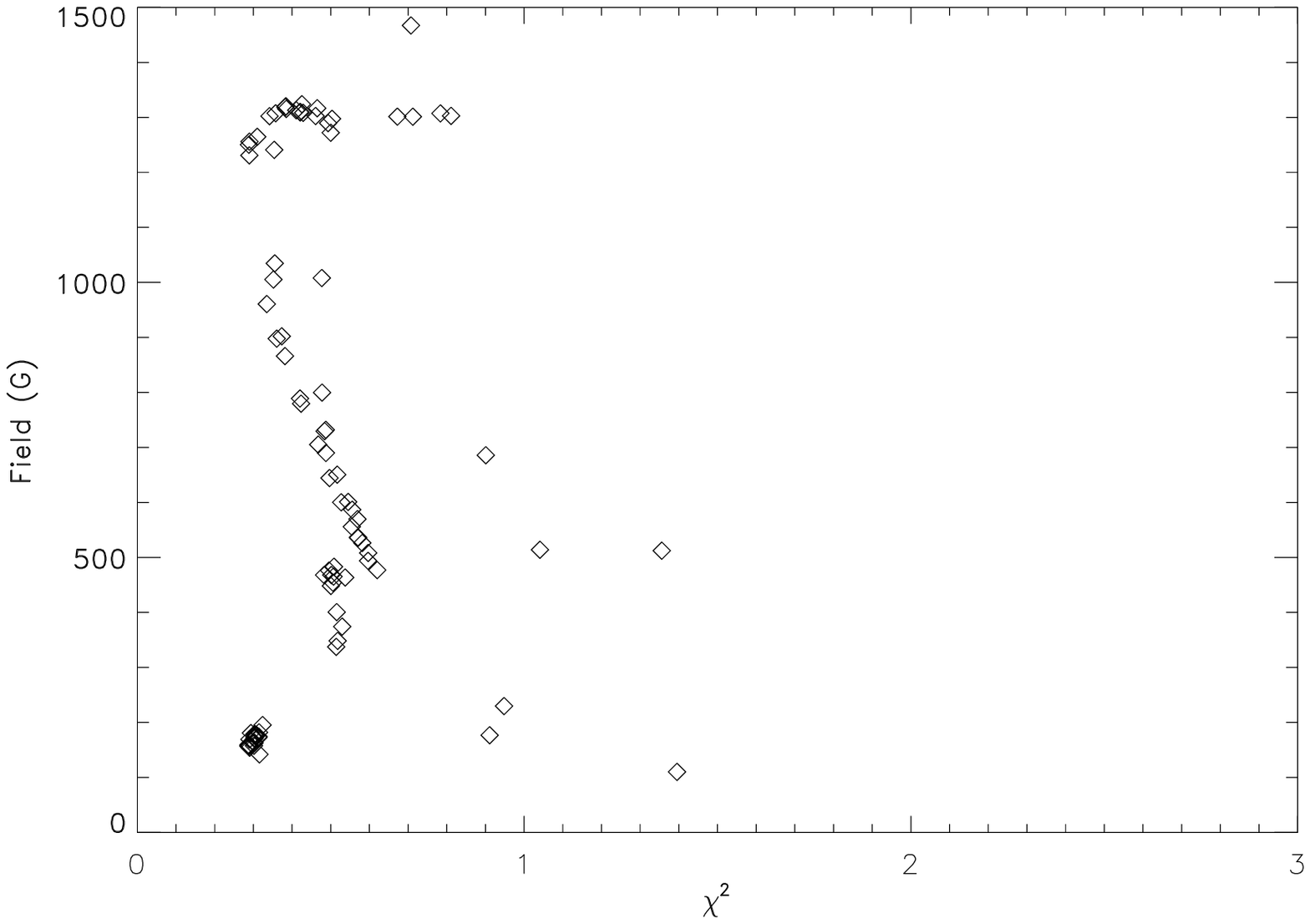}
\caption{Representation of the solutions from 100 different inversions
  of the 5250~\AA \, region with random initializations obtained using
  the Milne-Eddington code MELANIE. The field strength inferred by the
  inversion is plotted versus the quality of the fit, as measured by
  the merit function $\chi^2$.
\label{fig:uniqME5250}
}
\end{figure}

\begin{figure}
\epsscale{1}
\plotone{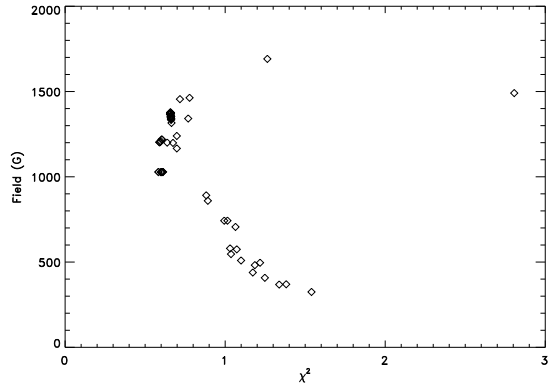}
\caption{Representation of the solutions from 100 different inversions
  of the 6302~\AA \, region with random initializations obtained using
  the Milne-Eddington code MELANIE. The field strength inferred by the
  inversion is plotted versus the quality of the fit, as measured by
  the merit function $\chi^2$.
\label{fig:uniqME6302}
}
\end{figure}

Figure~\ref{fig:NWa} depicts average profiles over a network
patch. The 5250 line ratio for this profile is 1.21. We started by
exploring the uniqueness of the magnetic field strength inferred with
the simplest algorithm, MELANIE. Each one of these average profiles
was inverted 100 times with random initializations. The results are
presented in Figures~\ref{fig:uniqME5250} and~\ref{fig:uniqME6302},
which show the values obtained versus the goodness of the fit, defined
as the merit function $\chi^2$, which in this work is defined as:
\begin{equation}
\label{eq:chisq}
\chi^2={1 \over N_p} \sum_{i=1}^{N_p}{ (I_i^{obs}-I_i^{syn})^2 \over
  \sigma_i^2} \, ,
\end{equation}
where $N_p$ is the number of wavelengths and $\sigma_i$ have been
taken to be 10$^{-3}$, so that a value of $\chi^2$=1 would represent
on average a good fit at the 10$^{-3}$ level. The average profiles
inverted here have a much lower noise (near 10$^{-4}$) and thus it
is some times possible to obtain $\chi^2$ smaller than 1.

The $\chi^2$ represented in the plots
is the one corresponding to the Stokes~$V$ profile only (although the
inversion codes consider both $I$ and $V$, but $I$ is consistently
  well reproduced and does not help to discriminate among the
  different solutions). Most of the
fits correspond to kG field, indicating that inversions of network
profiles are very likely to yield high field strengths. However, there
exists a very large spread of field strength values that provide
reasonably good fits to the observed data. This is especially true for
the 5250~\AA \, lines, for which it is possible to fit the
observations virtually equally well with fields either weaker than
500~G or stronger than 1~kG. In the case of 6302~\AA \, the best
solutions are packed around $\simeq$1.5~kG, although other solutions
of a few hecto-Gauss (hG) are only slightly worse than the best fit.

\begin{figure}
\epsscale{1}
\plotone{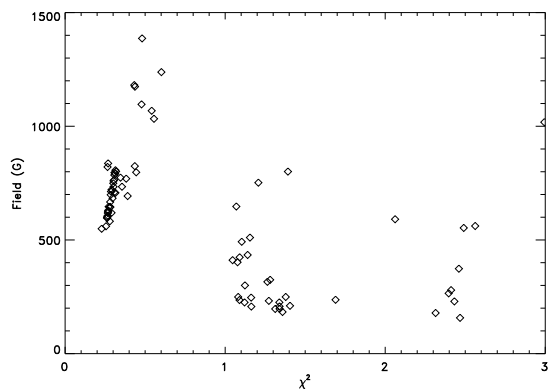}
\caption{Representation of the solutions from 100 different inversions
  of the 5250~\AA \, region with random initializations obtained using
  the LTE code LILIA. The field strength inferred by the
  inversion is plotted versus the quality of the fit, as measured by
  the merit function $\chi^2$.
\label{fig:uniqLI5250}
}
\end{figure}

\begin{figure}
\epsscale{1}
\plotone{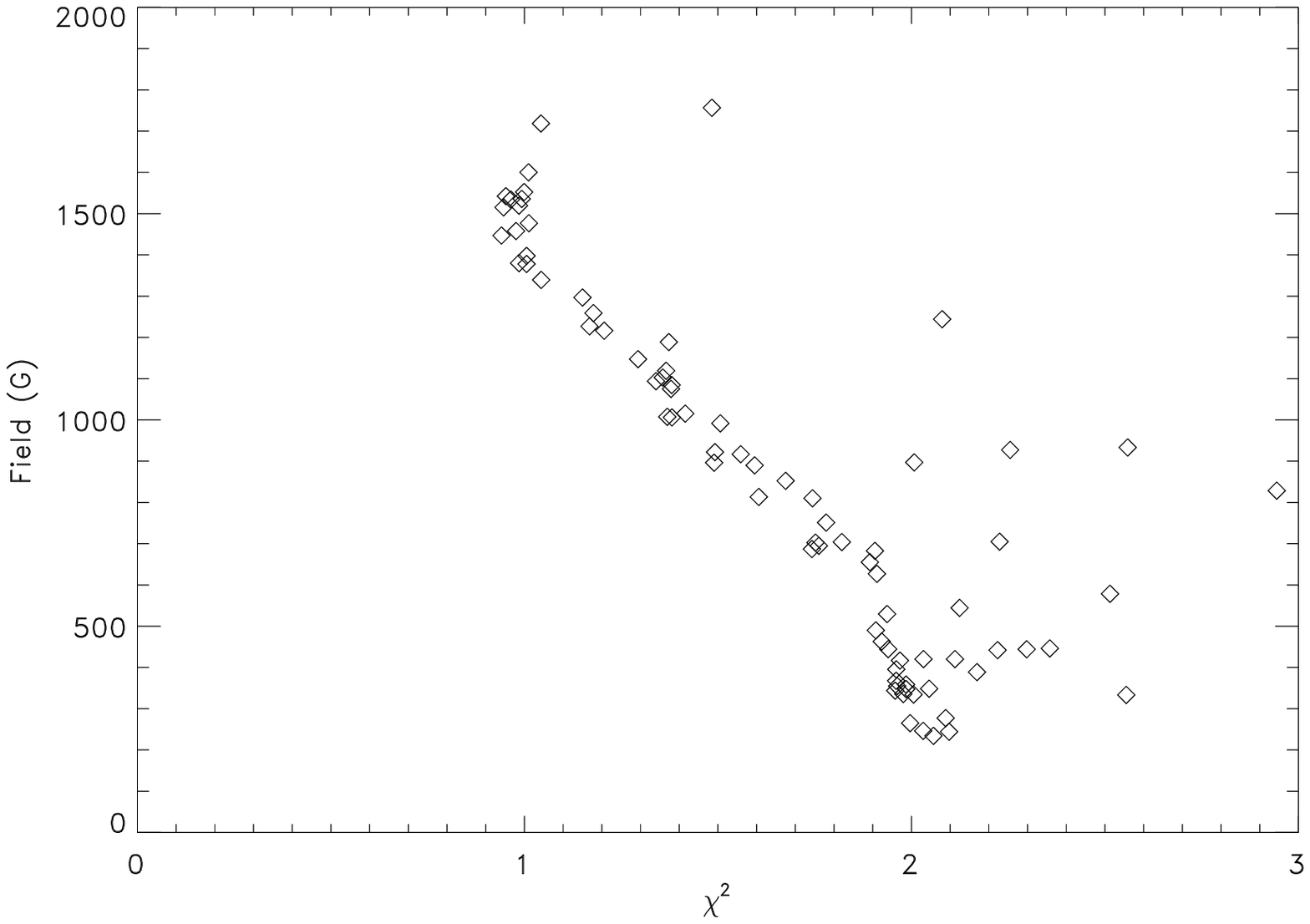}
\caption{Representation of the solutions from 100 different inversions
  of the 6302~\AA \, region with random initializations obtained using
  the LTE code LILIA. The field strength inferred by the
  inversion is plotted versus the quality of the fit, as measured by
  the merit function $\chi^2$.
\label{fig:uniqLI6302}
}
\end{figure}

\begin{figure*}
\epsscale{2}
\plotone{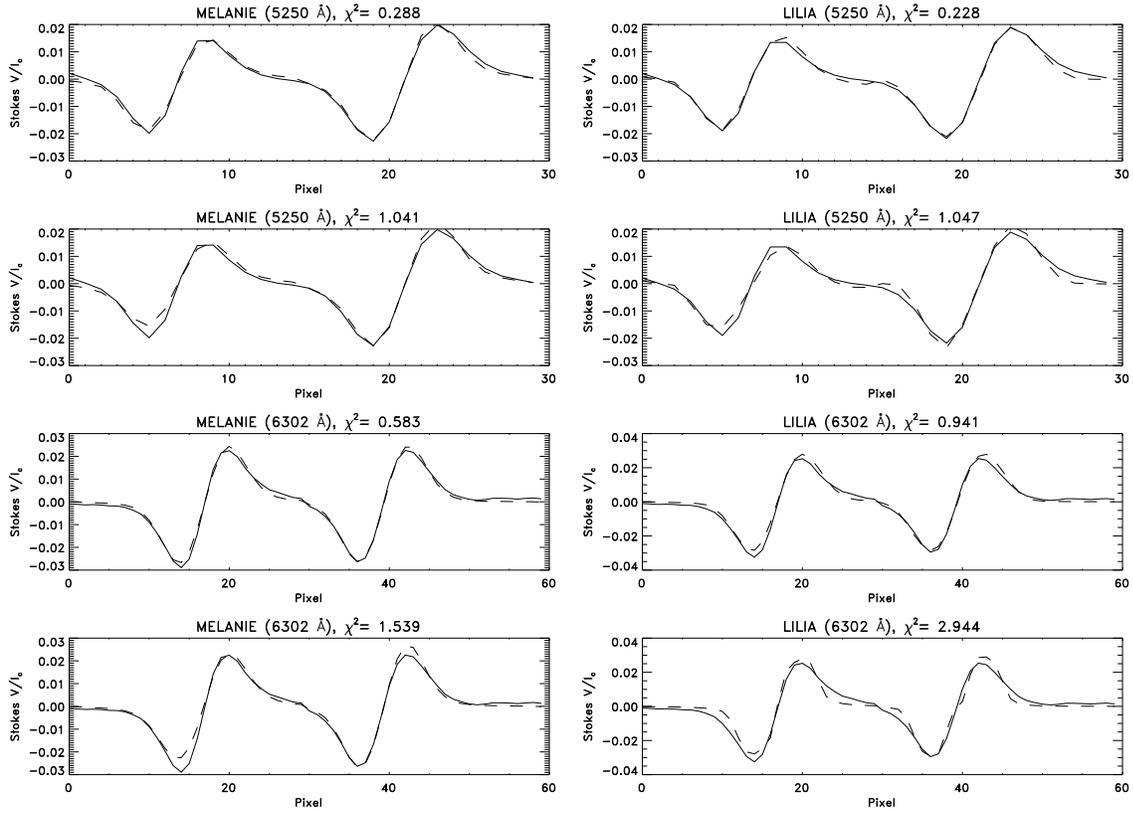}
\caption{Fits obtained with MELANIE (left) and
  LILIA (right) to the average network profile observed at 5250
  (upper four panels) and 6302~\AA \, (lower four panels). For each
  case we show the best fit (smaller $\chi^2$) and a not so good one.
\label{fig:fits}
}
\end{figure*}

It could be argued that Milne-Eddington inversions are too simplistic
to deal with network profiles, since they are known to exhibit fairly
strong asymmetries (both in area and in amplitude) that cannot be
reproduced by a Milne-Eddington model. With this consideration in
mind, we made a similar experiment using the LILIA and SIR codes. 
Figures~\ref{fig:uniqLI5250} and~\ref{fig:uniqLI6302} show the results
obtained with LILIA. The magnetic and non-magnetic atmospheres have
been forced to have the same thermodynamics in order to reduce the
possible degrees of freedom (on the downside, this introduces an
implicit assumption on the solar atmosphere). Again, the 6302~\AA \,
lines seem to yield somewhat more robust inversions.  The best fits
correspond to field strengths of approximately 1.5~kG, with weaker
fields delivering somewhat lower fit quality. The 5250~\AA \, lines
give nearly random results (although they tend to be clustered between
500 and 800~G there is a tail of good fits with up to almost
1400~G). Similar results are obtained using SIR. In 
order to give an idea of what the different $\chi^2$ values mean, we
present some of the fits in Fig~\ref{fig:fits}.

\begin{figure}
\epsscale{1}
\plotone{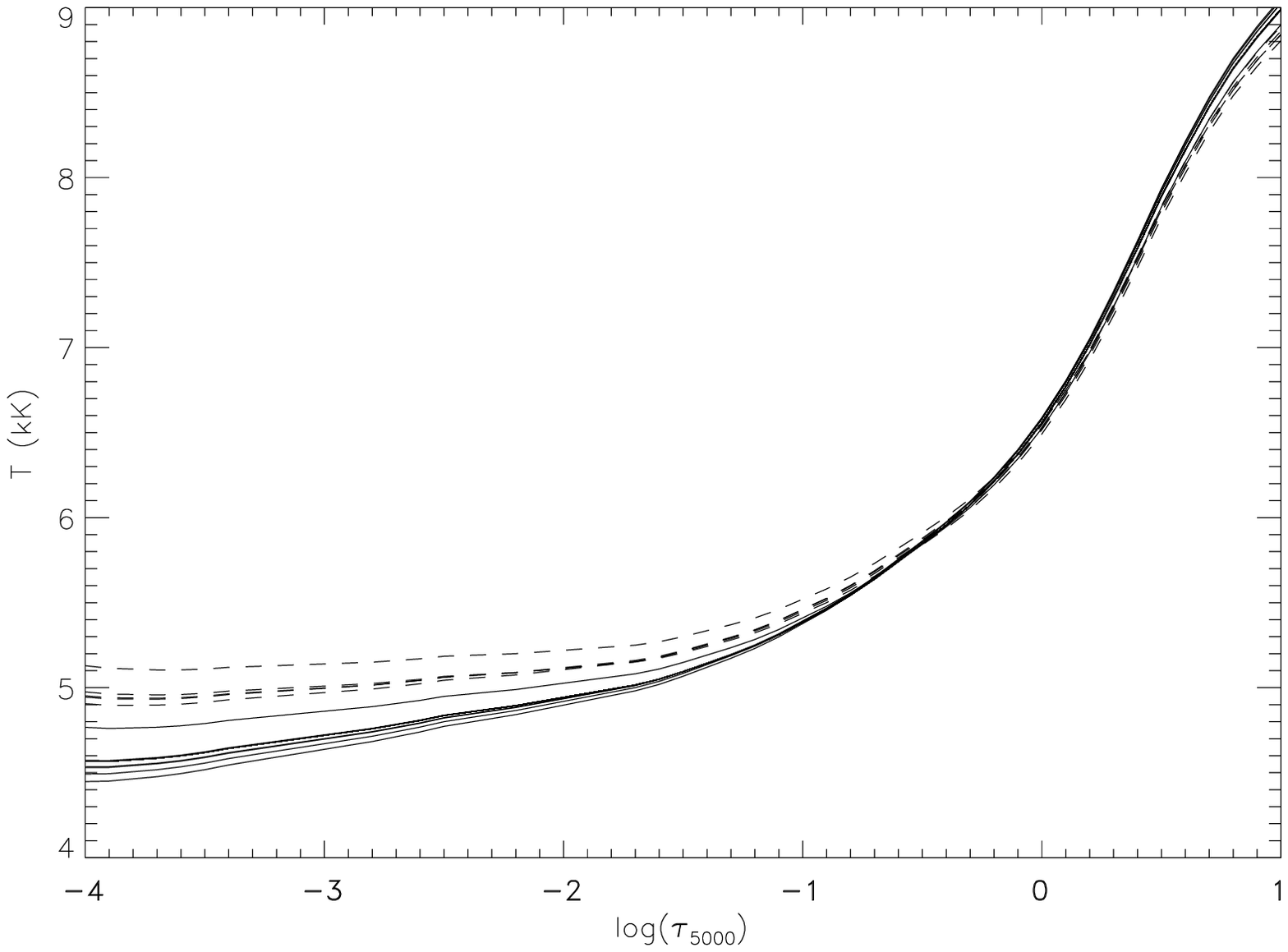}
\caption{Temperature stratification of the models obtained from the
  6302~\AA \, inversions with LILIA. Solid line: Models that include
  kG fields. Dashed line: Models with sub-kG field.
\label{fig:tlil6302}
}
\end{figure}

\begin{figure}
\epsscale{1}
\plotone{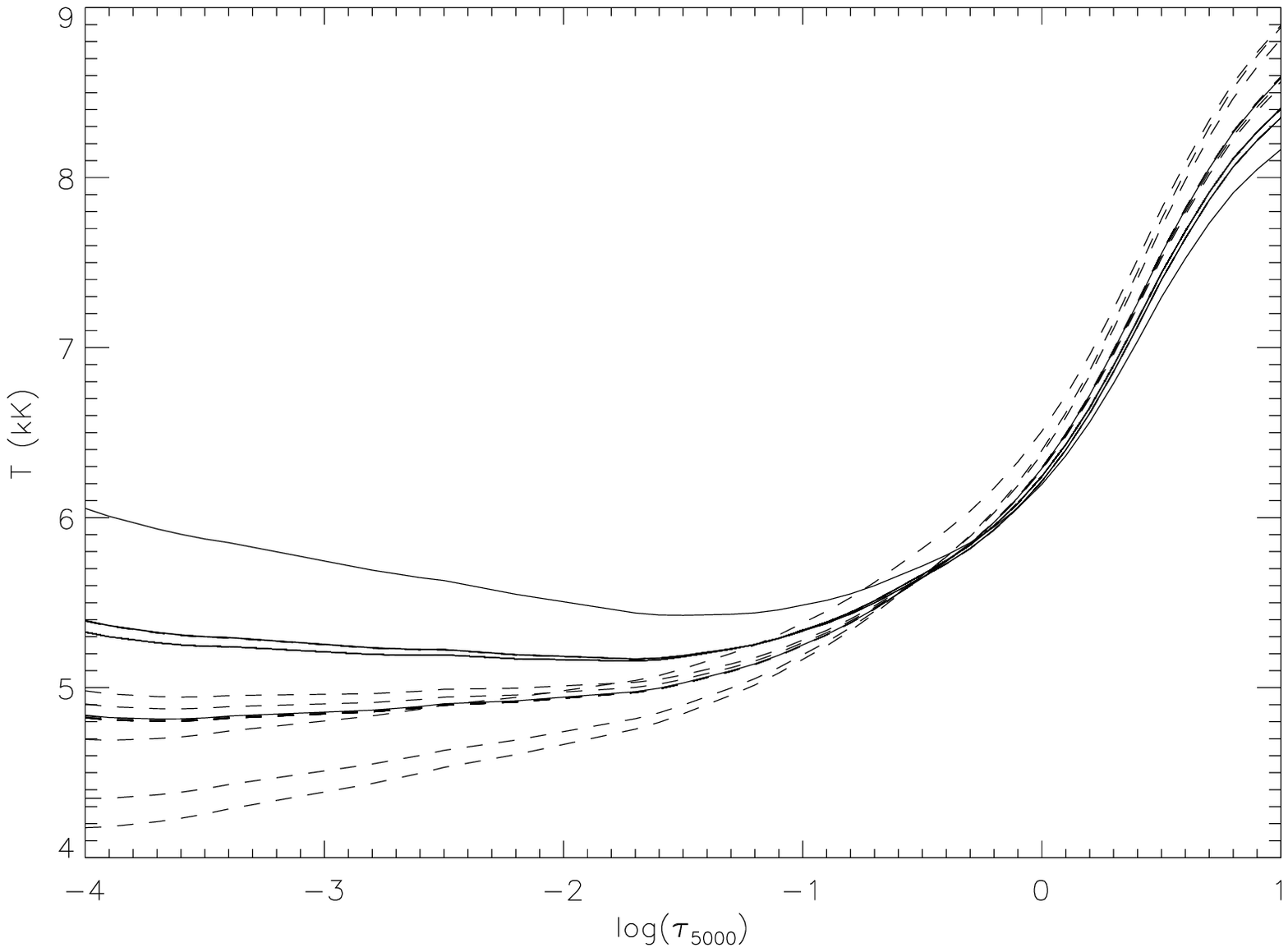}
\caption{Temperature stratification of the models obtained from the
  5250~\AA \, inversions with LILIA. Solid line: Models that include
  kG fields. Dashed line: Models with sub-kG field.
\label{fig:tlil5250}
}
\end{figure}

The thermal stratifications obtained in the 6302~\AA \, inversions are
relatively similar, although the weaker fields require a hotter upper
photosphere than the stronger fields (see Fig~\ref{fig:tlil6302}). On
the other hand, the 5250~\AA \, inversions do not exhibit a clear
correlation between the magnetic fields and temperature inferred
(Fig~\ref{fig:tlil5250}).

\begin{figure}
\epsscale{1}
\plotone{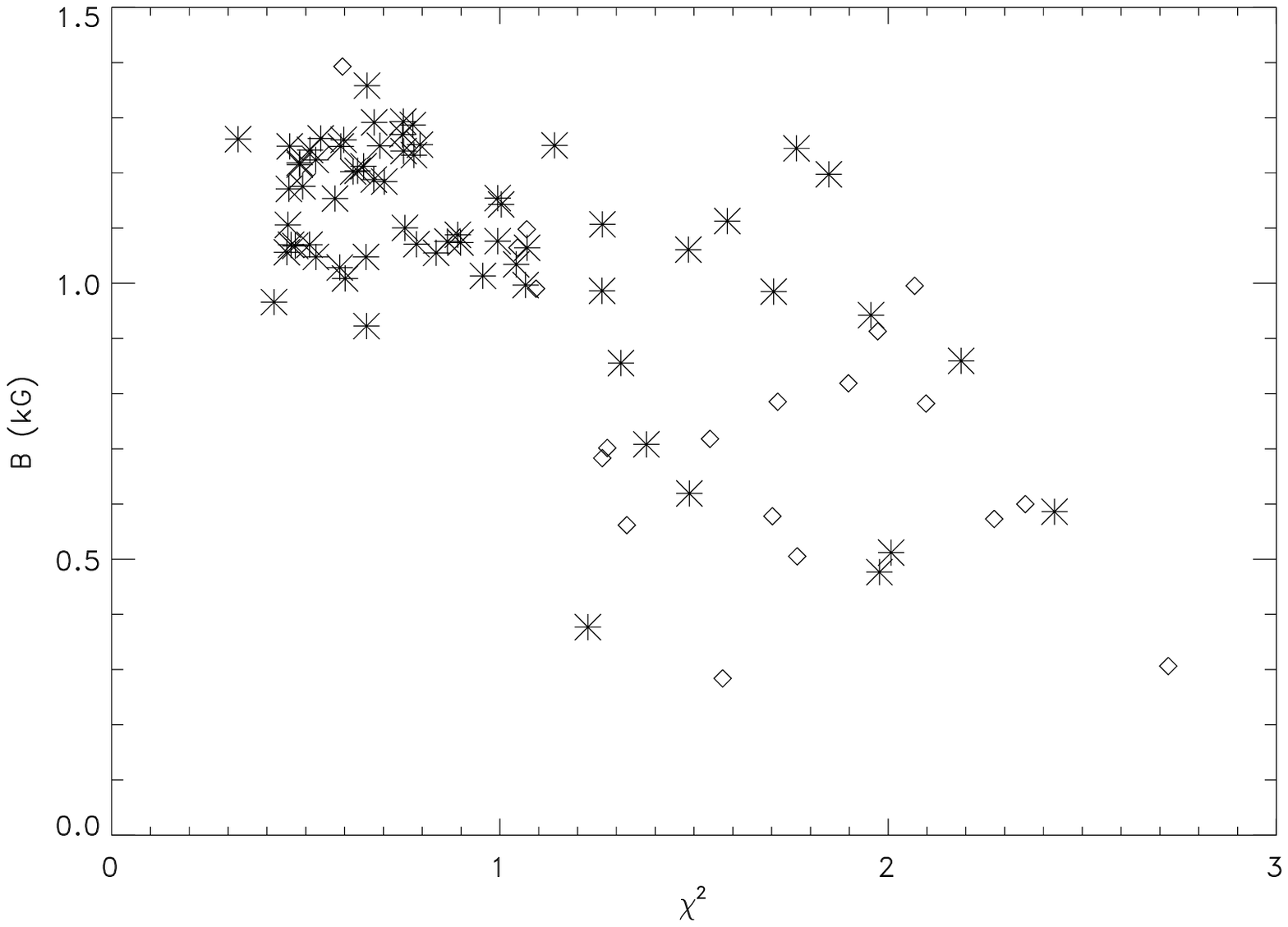}
\caption{Inversions of the 6302~\AA \, region. Average magnetic field
  in the lower photosphere of the MISMA component that harbors more
  flux, as a function of the quality of the fit $\chi^2$. Asterisks:
  Solutions where the temperature decreases outwards in the upper
  photosphere. Diamonds: Solutions where the temperature increases
  outwards in the upper photosphere.
\label{fig:misma6302}
}
\end{figure}

\begin{figure}
\epsscale{1}
\plotone{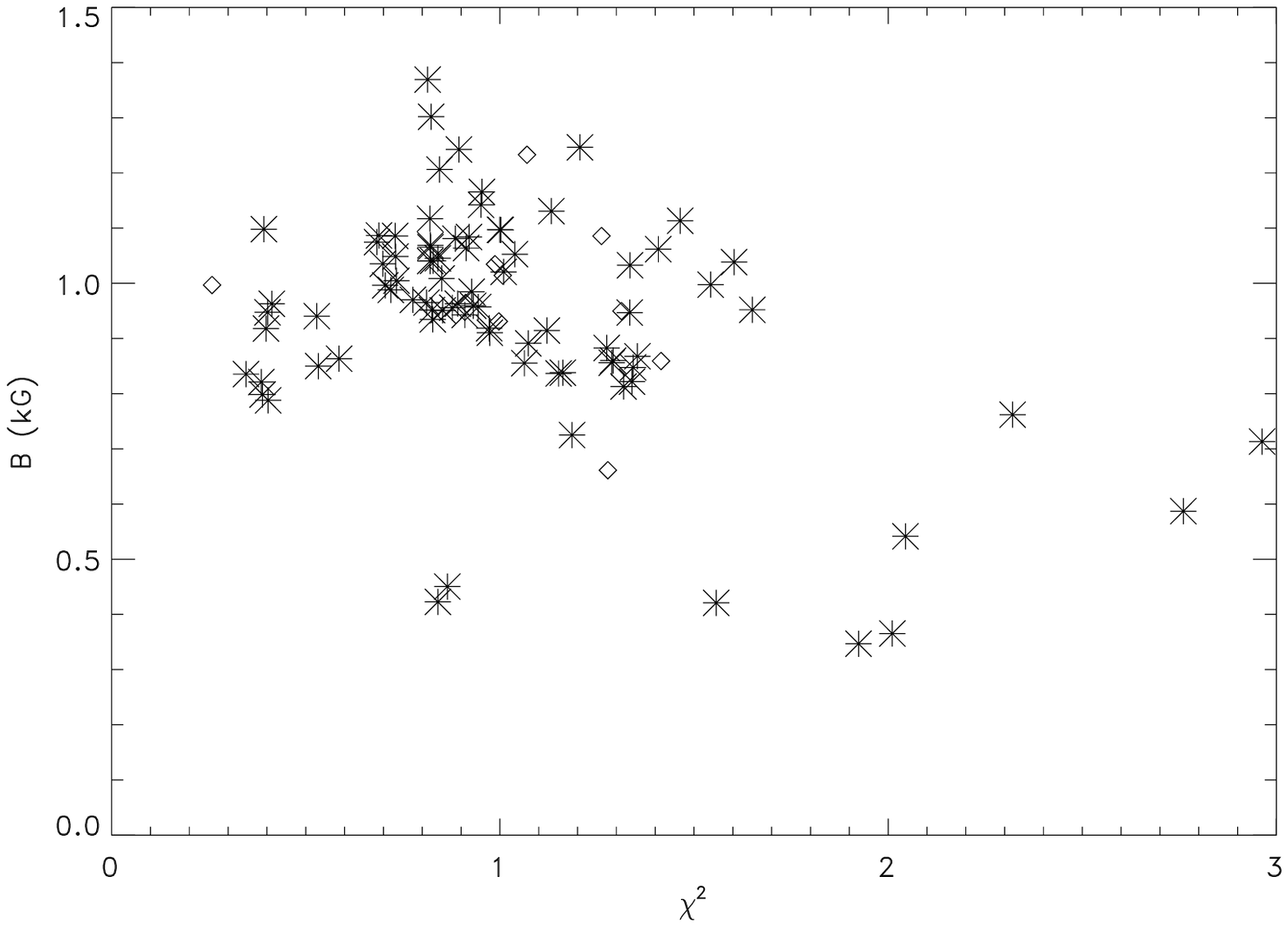}
\caption{Inversions of the 5250~\AA \, region. Average magnetic field
  in the lower photosphere of the MISMA component that harbors more
  flux, as a function of the quality of the fit $\chi^2$. Asterisks:
  Solutions where the temperature decreases outwards in the upper
  photosphere. Diamonds: Solutions where the temperature increases
  outwards in the upper photosphere.
\label{fig:misma5250}
}
\end{figure}

The MISMA code was also able to find good solutions with
either weak or strong fields. The smallest $\chi^2$ values
  correspond systematically to kG fields, but there are also
  some reasonably good fits ($\chi^2 \simeq 1$) obtained with weak
  fields of $\sim$500~G (Figs~\ref{fig:misma6302}
and~\ref{fig:misma5250}). However, we found that all the 
weak-field solutions for 6302~\AA \, have a temperature that increases
outwards in the upper photosphere (Fig~\ref{fig:misma6302}). This
might be useful to discriminate between the various solutions. The
5250~\AA \, lines, on the other hand, do not exhibit this behavior
(Fig~\ref{fig:misma5250}). 

In principle it would seem plausible to discard the models with
outward increasing temperature using physical arguments. This would
make us conclude from the 6302~\AA \, lines that the fields are
actually very strong, between $\sim$1.5 and~2.5~kG (it would not be
possible to draw similar conclusions from the 5250~\AA \, lines). In
any case, it would be desirable to have a less model-dependent
measurement that could be trusted regardless of what the thermal
stratification is. 

\begin{figure}
\epsscale{1}
\plotone{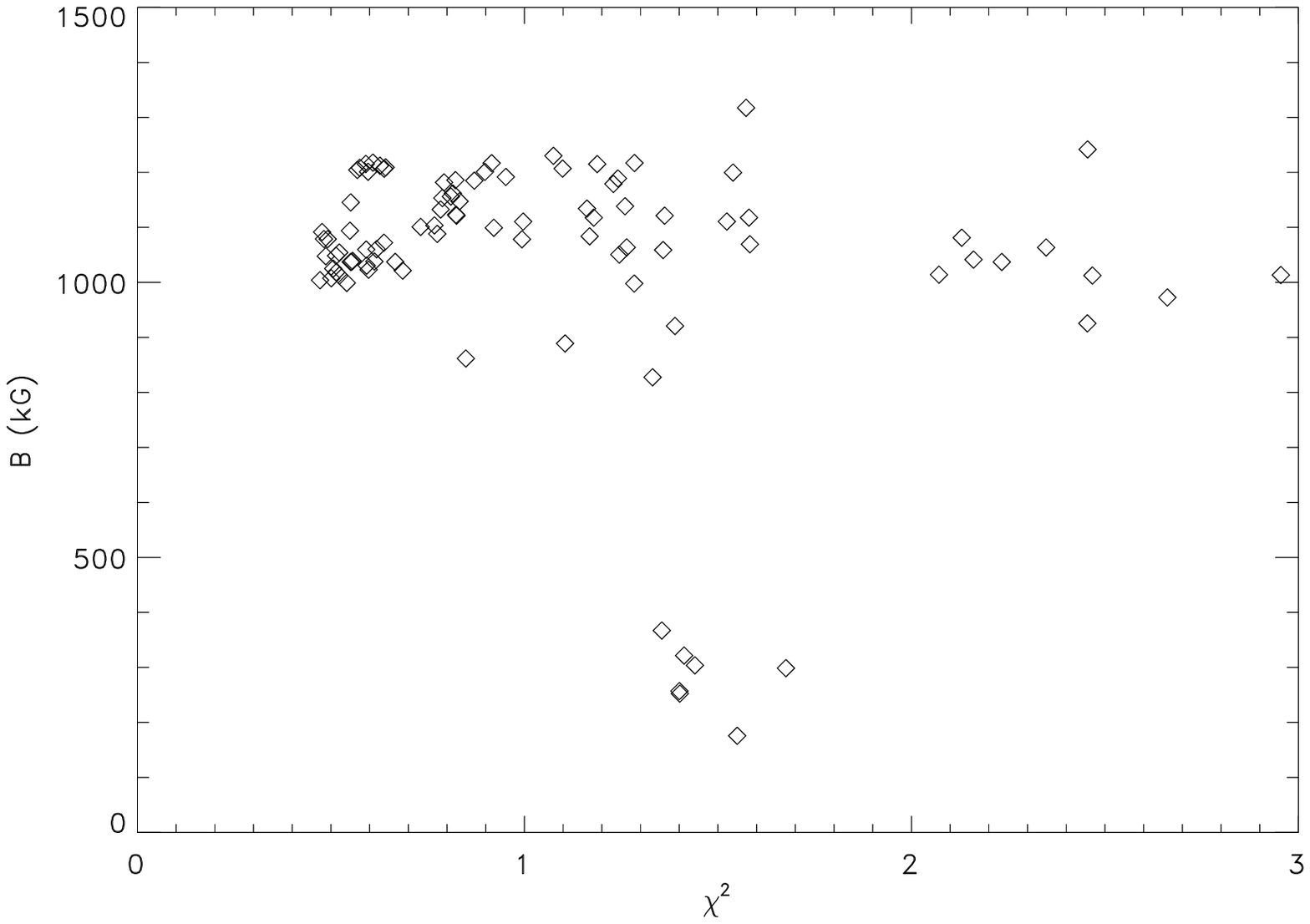}
\caption{Inversions of combined 5250 and 6302~\AA \, profiles. Average
  magnetic field in the lower photosphere of the MISMA component that
  harbors more flux, as a function of the quality of the fit $\chi^2$.
\label{fig:mismaboth}
}
\end{figure}

Interestingly, when we invert all four \ion{Fe}{1} lines
simultaneously, both at 5250 and 6302~\AA , the weak-field solutions
disappear from the low-$\chi^2$ region of the plot and the best
solutions gather between approximately 1 and 1.4~kG (see
Fig~\ref{fig:mismaboth}).

\citeN{SNARMS06} present a list of spectral lines with identical
excitation potentials and oscillator strengths. We decided to test one
of the most promising pairs, namely the \ion{Fe}{1} lines at 4122 and
9000~\AA. The choice was made based on their equivalent widths, Land\'
e factors and also being reasonably free of blends in the quiet solar
spectrum. Simulations similar to those in Figure~\ref{fig:calib}
showed that the line-ratio technique is still incapable of retrieving
an unambiguous field strength due to the presence of line
broadening. Also, even  if the {\it line} opacities are the same, the
continuum opacities are significantly different at such disparate
wavelengths, but in any case the line opacity is much stronger than
the background continuum where the Stokes~$V$ lobes are formed.

When we tested the robustness of inversion codes applied to this pair
of lines, we obtained extremely reliable results as described
below. Unfortunately we do not have observations of these two lines
and therefore resorted on synthetic profiles to perform the tests. We
used a 2-component reference model, where both components have the
thermal structure of HSRA. The magnetic comonent has an arbitrary
magnetic field and line-of-sight velocity. The field has a linear
gradient and goes from $\simeq$1.6~kG at the base of the photosphere
to 1~kG at continuum optical depth $\log(\tau_{5000})$=-4. The
velocity field goes from 2.5~km~s$^{-1}$ to 0.4~km~s$^{-1}$ at
$\log(\tau_{5000})$=-4. The magnetic component has a filling factor of
0.2. The reference macroturbulence is 3~km~s$^{-1}$, which is roughly
the value inferred from our observations. The spectra produced by
this model at 5250 and 6302~\AA \, are very similar to typical network
profiles. We considered the reference 4122 and 9000~\AA \, profiles as
simulated observations and inverted them with two different methods.

\begin{figure}
\epsscale{1}
\plotone{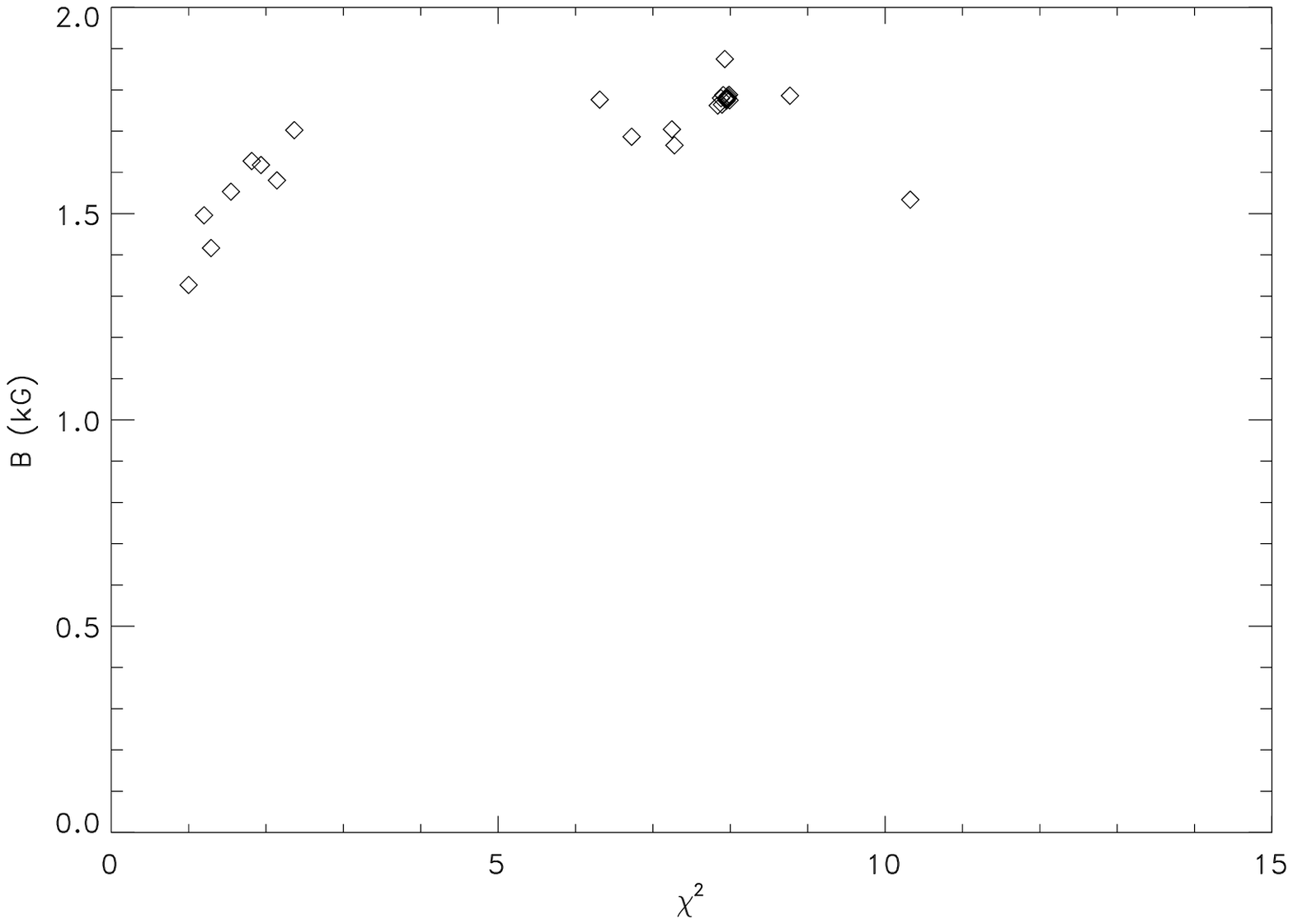}
\caption{Inversions of the proposed \ion{Fe}{1} 4122 and 9000~\AA \,
  profiles. Average magnetic field in the lower photosphere of the
  magnetic component as a function of the quality of the fit $\chi^2$.
\label{fig:lilia9000}
}
\end{figure}

In order to make the test as realistic as possible, we gave the
inversion code a somewhat erroneous non-magnetic
profile. Specifically, we multiplied the profile from the reference
non-magnetic component by a factor of 0.9 and shifted it 1 pixel
towards the red. The wavelength shift corresponds to roughly
0.36~km~s$^{-1}$ in both regions. This distorted non-magnetic profile
was given as input to the inversion codes. We inverted the reference
profiles using LILIA with 100 different initializations. Only
$\sim$30\% of the inversions converged to a reasonably low value of
$\chi^2$, with the results plotted in Figure~\ref{fig:lilia9000}. We
can see that the inversions are extremely consistent over a range of
$\chi^2$ much larger than in the previous cases. In fact, none of the
solutions are compatible with weak fields, suggesting that these lines
are much better at discriminating intrinsic field strengths.

A much more demanding verification for the diagnostic potential of
these new lines is to use a simpler scenario in the inversion than in
the synthesis of the reference profiles, incorporating typical
uncertainties in the calculation. After all, the real Sun will always
be more complex than our simplified physical models. Thus, inverting
the reference profiles with the Milne-Eddington code is an appropriate
test. For this experiment we not only supplied the same ``distorted''
non-magnetic profile as above, but we also introduced a systematic
error in the $\log(gf)$ of the lines. We forced the opacity of the
9000~\AA \, line to be 20\% lower than that of 4122 (instead of taking
them to be identical, as their tabulated values would
indicate). Again, we performed 100 different inversions of the
reference set of profiles with random initializations. In this case
the inversion results are astonishingly stable, with 98 out of the 100
inversions converging to a $\chi^2$ within 15\% of the best fit. The
single-valued magnetic field obtained for those 98 inversions has a
median of 1780~G with a standard deviation of only 3~G. The small
scatter does not reflect the systematic errors introduced by several
factors, including: a)the inability of the Milne-Eddington model to
reproduce the comparatively more complex referece profiles; b)the
artificial error introduced in the atomic parameters of the 9000~\AA
\, line; c)the distortion (scale and shift) of the non-magnetic
profile provided to the inversion code.

A final caveat with this new pair is that, even though the synthetic
atlas of \citeN{SNARMS06} indicates that the 9000~\AA \, line Stokes~$V$
profile is relatively free of blends, this still needs to be confirmed
by observations (there is a very prominent line nearby that may
complicate the analysis otherwise).

\section{Conclusions}
\label{sec:conc}

The ratio of Stokes~$V$ amplitudes at 5250 and 5247~\AA \, is a very
good indicator of the intrinsic field strength in the absence of line
broadening, e.g. due to turbulence. However, line broadening tends to
smear out spectral features and reduce the Stokes~$V$ amplitudes. This
reduction is not the same for both lines, depending on the profile
shape. If the broadening could somehow be held constant, one would
obtain a line-ratio calibration with very low scatter. However, if the
broadening is allowed to fluctuate, even with amplitudes as small as
1~km~s$^{-1}$, the scatter becomes very large.  Fluctuations in the
thermal conditions of the atmosphere further complicate the
analysis. This paper is not intended to question the historical merits
of the line-ratio technique, which led researchers to learn that
fields seen in the quiet Sun at low spatial resolution are mostly of
kG strength with small filling factors. However, it is important to
know its limitations. Otherwise, the interpretation of data such as
those in Figure~\ref{fig:mapratios} could be misleading. Before this
work, most of the authors were under the impression that measuring the
line ratio of the 5250~\AA \, lines would always provide an accurate
determination of the intrinsic field strength.

With very high-resolution observations, such as those expected from
the Advanced Technology Solar Telescope (ATST, \citeNP{KRK+03}) or the
Hinode satellite, there is some hope that most of the turbulent
velocity fields may be resolved. In that case, the turbulent
broadening would be negligible and the line-ratio technique would be
more robust. However, even with the highest possible spatial
resolution, velocity and temperature fluctuations along the line of
sight will still produce turbulent broadening.

From the study presented here we conclude that, away from active
region flux concentrations, it is not straightforward to measure
intrinsic field strengths from either 5250 or 6302~\AA \, observations
taken separately. Weak-flux internetwork observations would be
  even more challenging, as demonstrated recently by
  \citeN{MG07}. Surprisingly enough, the 6302~\AA \, pair of 
\ion{Fe}{1} lines is more robust than the 5250~\AA \, lines in the
sense that it is indeed possible to discriminate between weak and
strong field solutions if one is able to rule out a thermal
stratification with temperatures that increase outwards. Even so, this
is only possible when one employs an inversion code that has
sufficient MHD constrains (an example is the MISMA implementation used
here) to reduce the space of possible solutions.

The longitudinal flux density obtained from inversions of the 6302~\AA
\, lines is better determined than those obtained with 5250~\AA . This
happens regardless of the inversion method employed, although using a
code like LILIA provides better results than a simpler one such as
MELANIE. The best fits to average network profiles correspond to
strong kG fields, as one would expect.

An interesting conclusion of this study is that it is possible to
obtain reliable results by inverting simultaneous observations at both
5250 and 6302~\AA . Obviously this would be possible with relatively
sophisticated algorithms (e.g., LTE inversions) but not with simple
Milne-Eddington inversions.

The combination of two other \ion{Fe}{1} lines, namely those at 4122
and 9000~\AA , seems to provide a much more robust determination of
the quiet Sun magnetic fields. Unfortunately, these lines are very
distant in wavelength and few spectro-polarimeters are capable of
observing them simultaneously. Examples of instrument with this
capability are the currently operational SPINOR and THEMIS, as well as
the planned ATST and GREGOR. Depending on the evolution time scales
of the structures analyzed it may be possible for some other
instruments to observe the blue and red lines alternatively.

\acknowledgments
This work has been partially funded by the Spanish
Ministerio de Educaci\'on y Ciencia through project
AYA2004-05792


\end{document}